# Implications of mappings between ICD clinical diagnosis codes and Human Phenotype Ontology terms

Amelia LM Tan[1]*, PhD, Rafael S Gonçalves[2]*[†], PhD, William Yuan[1], PhD, Gabriel A Brat[1], MD, The Consortium for Clinical Characterization of COVID-19 by EHR (4CE), Robert Gentleman[2], PhD, Isaac S Kohane[1], MD, PhD

[1] Department of Biomedical Informatics, Harvard Medical School, Boston, MA
[2] Center for Computational Biomedicine, Harvard Medical School, Boston, MA

* Authors contributed equally to this paper.

[†] Corresponding author: Rafael_Goncalves@hms.harvard.edu

## Abstract

**Objective**
Integrating EHR data with other resources is essential in rare disease research due to low disease prevalence. Such integration is dependent on the alignment of ontologies used for data annotation. The International Classification of Diseases (ICD) is used to annotate clinical diagnoses; the Human Phenotype Ontology (HPO) to annotate phenotypes. Although these ontologies overlap in biomedical entities described, the extent to which they are interoperable is unknown. We investigate how well aligned these ontologies are and whether such alignments facilitate EHR data integration.

**Materials and Methods**
We conducted an empirical analysis of the coverage of mappings between ICD and HPO. We interpret this mapping coverage as a proxy for how easily clinical data can be integrated with research ontologies such as HPO. We quantify how exhaustively ICD codes are mapped to HPO by analyzing mappings in the UMLS Metathesaurus. We analyze the proportion of ICD codes mapped to HPO within a real-world EHR dataset.

**Results and Discussion**
Our analysis revealed that only 2.2% of ICD codes have direct mappings to HPO in UMLS. Within our EHR dataset, less than 50% of ICD codes have mappings to HPO terms. ICD codes that are used frequently in EHR data tend to have mappings to HPO; ICD codes that represent rarer medical conditions are seldom mapped.

**Conclusion**
We find that interoperability between ICD and HPO via UMLS is limited. While other mapping sources could be incorporated, there are no established conventions for what resources should be used to complement UMLS.

## Lay Summary

We present a thorough empirical analysis of the compatibility between ICD codes and HPO terms based on the UMLS Metathesaurus. ICD is used to annotate clinical diagnoses in EHR data, while HPO is used to annotate phenotypes in research databases. Bridging between the two artifacts is essential for health data integration and analysis. UMLS is a widely used source of cross-ontology mappings, and so it is important to quantitatively assess the extent to which ICD is mapped to HPO in the UMLS. The primary results from the paper include that a mere 2.2% of ICD codes in UMLS are directly linked to HPO. Furthermore, an analysis of our EHR dataset shows that less than half of the commonly used ICD codes can be mapped to HPO terms. Notably, commonly used ICD codes in EHR data tend to have corresponding mappings to HPO. In contrast, ICD codes representing rarer medical conditions are infrequently associated with HPO terms.

## Background and Significance

The analysis of clinical patient data together with experimental data is important in biomedical research and requires principled integration of these data. The ability to integrate real-world data is especially imperative in rare disease research, where disease prevalence is very low and therefore little data is available. The emergence of NLP tools has made it easier to structure free-text data from the EHR and to map (some) extracted data to the Unified Medical Language System (UMLS), which can then be used to traverse different ontologies [1]. Ontologies such as the Human Phenotype Ontology (HPO) [2] and WHO's International Classification of Diseases (ICD) [3] are often used to label patient data. The motivations or use cases for these two ontologies are distinct. In the US, ICD is used in EHRs primarily for reimbursement purposes, while HPO is used for biological characterization in research. These ontologies were developed independently, without coordination to date, and typically have limited interoperability between them. However, with the increasing use and usefulness of EHR data, a growing part of the "real-world data" characterization of patient populations and diseases and the interest in finer-grained characterizations for EHR and cohort databases, the uses of these two ontologies increasingly overlap. This has created an urgent motivation to improve interoperability between ICD and HPO.

While the two ontologies do fundamentally cover different scopes, the links between diseases and phenotypes are important, as they allow for the transformation of data from the EHR to the research context where phenotypes are essential in the characterization of diseases in research settings. While there are other ontologies that might be better suited for mapping from diseases in ICD codes to diseases in respective ontologies like MONDO, these are not as useful towards disease characterization and clinical diagnostics as some conditions do not have a disease label yet. There is a certain gray area between phenotypic feature and disease. Some diseases can still present themselves in HPO as features of other diseases (e.g. DM as a feature of Bardet Biedl syndrome). The extent of overlap is unknown. This exercise quantifies the overlap between these two domains of phenotypic features and disease.

To illustrate the lack of interoperability, consider the concept 'acute bronchitis', which is represented in ICD and HPO with different identifiers—J20 in ICD for 'acute bronchitis', and HP:0012388 in HPO. Without a mapping between these two symbols to establish that they are conceptually equivalent, and should be interpreted as the same thing, data that rely on only one of the codes cannot be easily cross-analyzed. Although there have been recent developments in NLP tools for ontology mapping purposes, there still lacks a consensus on how to use these tools across research groups to achieve accurate and reusable mappings between ontology terms. For this reason, many researchers rely on the UMLS Metathesaurus, which provides mappings between ontologies derived through expert curation, to attain partial interoperability among the ontologies in UMLS [4]. The UMLS is widely used for mapping across ontologies with approximately 50% of users using it for that purpose [5].

A survey that we did of recent research published in the last two years (see **Table 1**) demonstrates that researchers rely on UMLS as the primary source of mappings between HPO and ICD. Schofield et al. developed a method (reported in paper #1 in Table 1) to identify and gather disease-phenotype associations by leveraging the UMLS mappings of diseases represented in ICD to phenotypes from the HPO or Mammalian Phenotype (MP) ontologies [6]. In papers #2 to #4, authors used semi-automated mapping tools (specifically MetaMap, CLAMP, and cTakes) to identify HPO terms in free-text descriptions of phenotypes from EHR clinical notes. In the setting up of the Open Annotation for Rare Diseases (OARD)—a data resource containing annotations for rare-disease-related phenotypes, described in paper #4— the authors used the NLP tool cTakes to harmonize claim codes, lab procedures and clinical notes with UMLS CUIs before traversing to HPO and MONDO ontologies. In paper #5, the authors used existing mappings of ICD to HPO obtained from UMLS and BioPortal, and then a partial-logical mapping strategy that uses the HPO structure in order to obtain better mapping coverage of ICD to HPO terms. In paper #6, the authors applied automated string matching using the BERT NLP model followed by manual verification of disease name mappings. Paper #7 provides a Phecode–HPO mapping set generated by using mappings in UMLS, the Phecode map of ICD codes to Phecodes, and tool-generated mappings using different approaches. From this short survey, it is evident that recent research heavily depends on UMLS as the main source of mappings, hence the utility of quantifying the biases and implications of these mappings.

Table 1: Articles published in 2021-22 that used mappings between ICD and HPO.

| # | Title | Mapping Reference | Mapping Tool(s) | Year |
|---|---|---|---|---|
| 1 | Linking common human diseases to their phenotypes; development of a resource for human phenomics [6] | UMLS | None: employed newly introduced methods involving NLP + UMLS | 2021 |
| 2 | Clinical Phenotypic Spectrum of 4095 Individuals with Down Syndrome from Text Mining of Electronic Health Records [7] | UMLS | MetaMap | 2021 |
| 3 | Development of a phenotype ontology for autism spectrum disorder by natural language processing on electronic health records [8] | UMLS | CLAMP | 2022 |
| 4 | OARD: Open annotations for rare diseases and their phenotypes based on real-world data [9] | UMLS | cTakes | 2022 |
| 5 | Common genetic variation associated with Mendelian disease severity revealed through cryptic phenotype analysis [10] | BioPortal + UMLS | None: employed a partial logical mapping strategy | 2022 |
| 6 | Building a Knowledge Graph to Enable Precision Medicine [11] | UMLS | BERT | 2022 |

| 7 | Linking rare and common disease vocabularies by mapping between the human phenotype ontology and phecodes [12] | UMLS, PheMap | SORTA, string-matching, WikiMedMap | 2023 |

In this paper, we quantify the limitations and biases of using UMLS mappings of ICD to HPO and how they may affect research findings. Specifically, we set out to determine the coverage of mappings between ICD10-CM—the finer-grained variant of ICD10 used in the US—and HPO, as this indicates how much information in the EHR is "covered" when transiting between EHR coding (ICD) and research-oriented ontologies (such as HPO), and to discuss what that coverage implies for the (potential) integration and analysis of EHR data with experimental data.[1] We quantify how exhaustively the popular ICD classification has been mapped to HPO by analyzing mappings derived from the widely used source of mappings between biomedical ontologies—the UMLS Metathesaurus. The purpose of our work is not to specify new mappings, nor to identify erroneous mappings; rather we provide a synchronic analysis of the mappings that are publicly accessible and currently used to convert from ICD to HPO for such goals as rare disease analysis. We also discuss implications of those decisions for secondary research and analysis in the context of real-world EHR data from Beth Israel Deaconess Medical Center (BIDMC).

## Methods and Materials

### Ontologies analyzed

The HPO ontology logically and systematically describes phenotypic traits in human diseases. Many public disease knowledge databases, such as MedGen [13] and Orphanet [14], as well as consortiums like the Undiagnosed Disease Network (UDN) use HPO as the controlled vocabulary to annotate phenotypes for diseases. Various algorithms and tools also leverage HPO to support phenotype-based differential diagnostics, gene-disease discovery, and genomic diagnostics, among others. For example, the Phen2Gene tool uses a database of weighted and ranked gene lists for every HPO term and then, given a patient-specific list of HPO terms, the tool calculates a prioritized gene list based on a probabilistic model and outputs gene-disease relationships [15].

Real-world patient data from hospitals and health insurance claims data, on the other hand, are often stored using codes from WHO's International Classification of Diseases (ICD), which is primarily designed and used as a billing instrument. There is a vast amount of patient data encoded using these ICD annotations, including signs and symptoms, diseases, external causes of injury or diseases and abnormal findings [16]. Currently the most widely used version of ICD in healthcare settings throughout the US is ICD10-CM, which replaces the older ICD9-CM.

### Sources of ontology mappings

As we mentioned, UMLS appears to be the most popular and comprehensive source of mappings. However there are other potential sources of mappings that researchers can consider when integrating data annotated with different ontologies. BioMappings [17] provides a community-contributed table of mappings—some verified by humans, while others simply predicted using software tools. At the time of writing, there were zero mappings between ICD and HPO in the BioMappings repository. Ontologies themselves often include mappings between the terms they define and terms in other ontologies or databases (so-called "database cross-references"). The ICD classification itself does not contain any such mappings, however HPO does. There are 39 mappings between HPO and ICD terms, with 20 of those only present in HPO, and not in UMLS. These "missing" mappings could be due to UMLS not being up-to-date with the latest release of HPO. We also extracted 632 ICD mappings from the MONDO disease ontology, and which have an HPO mapping. We determined that all 632 mappings are already contained in UMLS.

---

[1] From here on out, any mention of "ICD" is meant as a shorthand for "ICD10-CM."

**UMLS as a current mapping source**

The UMLS Metathesaurus unifies concepts across ontologies via an assignment of concept unique identifiers (CUI). A CUI is assigned to each unique collection of terms that are conceptually equivalent. For our study, we used the ICD CUIs and searched for HPO terms with identical CUIs in the May 2022 version of the UMLS tables. More specifically, after downloading the UMLS archive (umls-2022AA-mrconso.zip), we use in our analysis the file MRCONSO.RRF (of 2GB in size) contained therein—this is a table containing all UMLS CUIs and their associated labels, synonyms, and mappings to different ontologies. The table comes without column names; in our analysis we primarily use columns 1 (CUI), column 12 (SAB—abbreviated source ontology), column 13 (CODE—ontology term identifier), and column 15 (STR—string label). For example, the following sample of the UMLS table contains two mappings: (1) between ICD10CM code K85 and HPO term HP:0001735, and (2) between K85.9 and HP:0001735, because they share the same CUI. When a CUI has both an ICD code *(I)* and an HPO term *(H)* associated with it, we say that *I* is mapped to *H*.

```
CUI | SAB | CODE | STR
C0001339 | ICD10CM | K85 | Acute pancreatitis
C0001339 | ICD10CM | K85.9 | Acute pancreatitis, unspecified
C0001339 | HPO | HP:0001735 | Acute pancreatitis
C0001339 | HPO | HP:0001735 | Acute pancreatic inflammation
```

In 2014, Bodenreider et al. analyzed the coverage of disease phenotypes in standard biomedical ontologies to determine which phenotypes have a mapping in UMLS, and to which specific ontology(ies) in UMLS they map to [18]. A mapping to UMLS was found for 54% of disease phenotypes, with the best coverage by a single ontology being provided by SNOMED CT, which covers 30% of phenotypes in HPO. According to the study, at the time ICD10-CM provided coverage for only 15% of HPO phenotype terms [19].

**Mapping coverage of codes with different levels of usage in BIDMC data**

Our BIDMC EHR dataset was extracted in February 2022, and is a subset of the data submitted to the 4CE Consortium [20]. To understand the mapping coverage of codes with different levels of usage, we split the ICD codes into three usage groups. Codes that are used across more than 1% of patients are referred to as common codes; Codes that are used in 1-0.1% of patients are categorized as infrequently used, and codes that are used in less than 0.1% of patients are grouped as rare ICD codes. For each of these groups, we computed the proportion of mapped and unmapped codes as well as the proportion of codes that fell into the "others" category. Codes in the "others" category are those found in the ICD code list in our EHR dataset, but which did not match existing ICD10-CM codes. Throughout our analyses we stratified the setting of care into three groups: ICU patients, admitted patients and outpatients. The rationale for this stratification is that the setting of care might confound the mapping coverage, since codes used in outpatients could be very different from codes used in ICU patients. Furthermore, use cases for different care settings may differ, so it would be informative to characterize mapping coverage across these patient groups. For example, researchers working with ICU data who want to convert their ICD codes to HPO terms might face a different challenge compared to researchers working with outpatients when performing the same task.

### Mappability of the most used ICD codes

To survey the mappability of the most used ICD codes, we retrieved the 10 ICD codes with the highest number of patients counts without an associated HPO term and those with a mapped HPO term. For each of these codes we extracted the number of times the codes were used (counts) and the number of patients that have been assigned these codes (Patient Counts). For those with a matched HPO term, the matching HPO term and label are also presented.

### Mappability of ICD code categories by top-level ICD categories

We stratified our next analyses by top-level ICD categories to better understand if certain disease groups are better mapped. To further differentiate the codes, we split them into two groups: one for less-used codes which are used in less than 100 patients (inclusive), and codes that are more commonly used in more than 100 patients.

We then calculate the PatientCoverage for each of these groups of ICD codes. The purpose of calculating PatientCoverage was to quantify the proportion of patients with ICD codes from each ICD category that have HPO mappings. To calculate this, we take the summation of the total number of patients with mapped ICD codes as a proportion of the total number of patients with codes assigned in the ICD category. For example, consider the codes used in <100 patients from the category "Q: Congenital malformations, deformations and chromosomal abnormalities"— there are 309 codes used a total of 3506 times. Of these usage counts, 1260 of them involve mapped terms and hence the proportion mapped for this ICD category in terms of usage counts is 1260/3506 resulting in a patient coverage proportion of 0.359.

$$PatientCoverage = \frac{\Sigma PatientCounts\ for\ matched\ codes}{\Sigma PatientCounts\ for\ all\ codes\ in\ category}$$

(1)

## Results

### Dictionary level mapping

We first collated the dictionary-level mapping of all ICD codes to HPO, which is composed of all mappings between ICD codes and HPO terms in UMLS. The results from this analysis show that 2.2% of ICD codes are mappable to HPO via the UMLS Metathesaurus (Table 2). Although the proportion mappable is low, the distribution of mapped codes within real-world hospital ICD usage would provide a practical estimate of the consequences of the sparse mapping.

Table 2. Mappings between ICD and HPO in UMLS. ICD10-CM is the coding system used at BIDMC.

| Ontology | # Codes | # *Unique* HPO codes Mapped to ICD | % HPO codes Mapped to ICD | # *Unique* ICD codes Mapped to HPO | % ICD Mapped to HPO (#unique codes mapped/total ICD codes) |
|---|---|---|---|---|---|
| HPO | 16 366 | 1870 | 11.4 % | - | - |
| ICD10-CM | 95 847 | - | - | 2066 | 2.2 % |

**Mapping coverage of codes with different levels of usage**

Therefore, we investigated the mappability of codes across different levels of usage with the categories for commonly used ICD codes (>1%), codes that are infrequently used (0.1-1%) and codes that are rarely used (<0.1%). For the commonly used codes (>1%), the ICU patients have highest proportion of mapped codes (33.3%) as compared to the admitted patients and outpatients which have 24.4% and 25.8% of the ICD codes matched to a corresponding matched HPO term respectively (Figure 1). ICU patients also show a higher proportion mapped for the infrequently used ICD codes as compared to the admitted and outpatient cohorts possibly due to better-defined codes that are used in the ICU (Figure 1). Surprisingly, the outpatient group also does marginally better than the admitted patients for the commonly used codes. Overall, although less than half of the codes used are mappable to HPO via UMLS, it is reassuring that codes that are used more frequently tend to have a higher proportion mapped to HPO.

**Mappability of the most used ICD codes**

To survey the mappability of the most used ICD codes, we retrieved the 10 ICD codes with the highest number of patients counts with (Table 4) and without an associated HPO term (Table 3). For example, Table 3 presents some ICD codes that are used in roughly 2000 patients or more, and which do not have a corresponding HPO term, hence representing the greatest loss of information when converting between ICD and HPO terms. Of the codes that are not mapped, there were many (more than 7 out of 10 codes) that were from the ICD category 'Z' that covers codes that are "factors influencing health status and contact with health services". These codes mainly cover encounters with the health system and hence the absence of mappings is expected since the scope of HPO is to describe phenotypes. While we could choose not to interpret these as missing mappings, as we expect them to be missing, we point out that there are some Z codes in ICD that are mapped to HPO—for example, Z67.1 (Type A blood) is mapped to HP:0032370 (Blood group A). So to emulate the process of converting EHR data, we did not remove Z codes from our analysis since we want to characterize the overall mappability of ICD to HPO. Also, this category of codes only accounted for 5.7% of all codes used in admitted patients and does not significantly affect overall mappability.

The other codes which are poorly mapped range from ones like "anxiety NOS" and "major depressive disorder, single episode, unspecified", "acute posthemorrhagic anemia", to "acute respiratory failure with hypoxia" which are common diseases with multiple possible causes. Of these unmapped codes, for example, "acute respiratory failure with hypoxia", a manual lookup on the HPO website using keywords "respiratory failure" yielded a close match "Respiratory failure HP:0002878" which could potentially be mapped to the ICD code and hence better capturing the biology of the 2514 patients associated with the ICD code "J96.01 acute respiratory failure with hypoxia". This suggests that such trivial manual improvements to mapping bring about a very favorable effort to pay-off ratio in bringing about a much-improved patient representation. If capturing the overall population diagnosis is the goal, then these codes represent the lowest hanging fruits for the highest improvement in patient diagnosis representation. It also points the way to effective use of machine learning techniques for such mappings.

The most frequently used codes which are mappable to HPO terms via UMLS represent the common diagnoses that are most well-represented. When studying common diagnoses one can reasonably assume that the mapping between ICD and HPO would sufficiently capture the prevalence of those diagnoses in the EHR data. However, it should be cautioned that studies looking into diagnoses across diseases would have an overrepresentation of diagnoses in Table 4 since they are better mapped than those in Table 3.

Table 3: Table showing the 10 most frequently used codes in Admitted patients, ICU patients, and outpatients who are not admitted that do not have a matched HPO term.

| Cohort | Counts | Patient Counts | CUI | ICD ID | Label |
|---|---|---|---|---|---|
| Admitted | 15543 | 11671 | C5539297 | Z20.822 | Contact with and (suspected) exposure to COVID-19 |
| | 12544 | 7184 | C0085580 | Z87.891 | Personal history of nicotine dependence |
| | 11625 | 6659 | C2911355 | Z23 | Encounter for immunization |
| | 6283 | 6219 | C0341102 | Z37.0 | Single live birth |
| | 7006 | 5750 | C2910668 | Z20.828 | Contact with and (suspected) exposure to other viral communicable diseases |
| | 10348 | 5620 | C1313895 | F41.9 | Anxiety NOS |
| | 6721 | 5155 | C0022660 | Z11.59 | Encounter for screening for other viral diseases |
| | 8740 | 5042 | C2910658 | Z00.00 | Encounter for adult health check-up NOS |
| | 8073 | 4467 | C0003469 | F32.9 | Major depressive disorder, single episode, unspecified |
| | 11461 | 4405 | C2910579 | Z79.01 | Long term (current) use of anticoagulants |
| ICU | 5305 | 3506 | C0085580 | Z20.822 | Contact with and (suspected) exposure to COVID-19 |
| | 6029 | 2946 | C5539297 | Z87.891 | Personal history of nicotine dependence |
| | 3441 | 2845 | C0022660 | D62 | Acute posthemorrhagic anemia |
| | 3052 | 2514 | C2911355 | J96.01 | Acute respiratory failure with hypoxia |
| | 2831 | 2149 | C0154298 | Z20.828 | Contact with and (suspected) exposure to other viral communicable diseases |
| | 5752 | 2078 | C0341102 | Z79.01 | Long term (current) use of anticoagulants |
| | 2512 | 2010 | C2977065 | Z66 | DNR status |
| | 3910 | 1960 | C2882161 | F41.9 | Anxiety NOS |
| | 2733 | 1897 | C2910658 | Z11.59 | Encounter for screening for other viral diseases |
| | 3769 | 1828 | C2911178 | Z00.00 | Encounter for adult health check-up NOS |
| Outpatients | 116491 | 78109 | C2910579 | Z11.59 | Encounter for screening for other viral diseases |
| | 57453 | 37259 | C5203670 | U07.1 | COVID-19 |
| | 52548 | 33996 | C2910447 | Z00.00 | Encounter for adult health check-up NOS |
| | 39471 | 30929 | C2910668 | Z20.822 | Contact with and (suspected) exposure to COVID-19 |
| | 57339 | 29087 | C0085580 | Z23 | Encounter for immunization |
| | 26199 | 20805 | C5539297 | Z20.828 | Contact with and (suspected) exposure to other viral communicable diseases |
| | 24703 | 18863 | C5539296 | Z11.52 | Encounter for screening for COVID-19 |
| | 28360 | 13064 | C2910658 | F41.9 | Anxiety NOS |
| | 16779 | 12076 | C0341102 | Z00.01 | Encounter for general adult medical examination with abnormal findings |
| | 18014 | 11573 | C2910448 | Z87.891 | Personal history of nicotine dependence |

Table 4: Table showing the 10 most frequently used codes in admitted patients, ICU patients, and outpatients that are not admitted that have a matched HPO term.

| Cohort | Counts | Patient Counts | ICD ID | ICD Label | HPO | HPO Label |
|---|---|---|---|---|---|---|
| Admitted | 36074 | 11273 | I10 | high blood pressure | HP:0000822 | High blood pressure |
| | 23666 | 10802 | E78.5 | Hyperlipidemia, unspecified | HP:0003077 | Hyperlipidemia |
| | 13486 | 6791 | K21.9 | Esophageal reflux NOS | HP:0002020 | Gastroesophageal reflux disease |
| | 10538 | 6010 | N17.9 | Acute kidney failure, unspecified | HP:0001919 | Acute kidney failure |
| | 14376 | 4954 | I25.10 | Atherosclerotic heart disease NOS | HP:0001677 | Coronary atherosclerosis |
| | 10429 | 4834 | D64.9 | Anemia, unspecified | HP:0001903 | Anemia |
| | 7986 | 3908 | E66.9 | Obesity NOS | HP:0001513 | Obesity |
| | 5464 | 3530 | E87.1 | Sodium [Na] deficiency | HP:0002902 | Hyponatremia |
| | 4358 | 3386 | E87.2 | Acidosis | HP:0001941 | Acidosis |
| | 9044 | 3375 | E03.9 | Hypothyroidism, unspecified | HP:0000821 | Hypothyroidism |
| ICU | 10776 | 4494 | E78.5 | Hyperlipidemia, unspecified | HP:0003077 | Hyperlipidemia |
| | 14953 | 4431 | I10 | high blood pressure | HP:0000822 | High blood pressure |
| | 5596 | 3045 | N17.9 | Acute kidney failure, unspecified | HP:0001919 | Acute kidney failure |
| | 6049 | 2655 | K21.9 | Esophageal reflux NOS | HP:0002020 | Gastroesophageal reflux disease |
| | 7573 | 2422 | I25.10 | Atherosclerotic heart disease NOS | HP:0001677 | Coronary atherosclerosis |
| | 2938 | 2313 | E87.2 | Acidosis | HP:0001941 | Acidosis |
| | 4597 | 2032 | D64.9 | Anemia, unspecified | HP:0001903 | Anemia |
| | 3238 | 1997 | E87.1 | Sodium [Na] deficiency | HP:0002902 | Hyponatremia |
| | 3187 | 1398 | E66.9 | Obesity NOS | HP:0001513 | Obesity |
| | 1711 | 1385 | I95.9 | Hypotension, unspecified | HP:0002615 | Hypotension |
| Outpatients | 85112 | 23465 | I10 | high blood pressure | HP:0000822 | High blood pressure |
| | 44506 | 18526 | E78.5 | Hyperlipidemia, unspecified | HP:0003077 | Hyperlipidemia |
| | 30216 | 14182 | K21.9 | Esophageal reflux NOS | HP:0002020 | Gastroesophageal reflux disease |
| | 15559 | 9204 | R05 | Cough | HP:0012735 | Coughing |
| | 19115 | 8927 | E66.9 | Obesity NOS | HP:0001513 | Obesity |
| | 13624 | 7858 | R07.9 | Chest pain, unspecified | HP:0100749 | Chest pain |
| | 13525 | 7396 | R53.83 | Fatigue NOS | HP:0012378 | Fatigue |
| | 18379 | 6982 | G47.33 | Obstructive sleep apnea (adult) (pediatric) | HP:0002870 | Obstructive sleep apnea |
| | 9565 | 6947 | R51.9 | Headache, unspecified | HP:0002315 | Headache |
| | 13015 | 6575 | R06.02 | Shortness of breath | HP:0002094 | Dyspnea |

**Mappability of ICD code categories by top-level ICD categories**

As the top-level ICD categories represent different major disease groups (e.g., category I is for diseases of the circulatory system, category J is for diseases of the respiratory system), the next analysis is stratified by these categories to better understand any biases in the mapping of these disease groups (Figure 2). The outcome from this analysis shows that while codes that are used in less than 100 patients generally have a lower proportion mapped of less than 40% across all ICD categories, codes that are used in >100 patients have a much higher representation in the proportion of patient with a mapped code although the patient coverage does vary substantially depending on the ICD category (Figure 2). We conducted the same analysis of outpatient and ICU patient data, and the results are similar and in line with those of admitted patients that we just discussed (results not shown).

# Discussion

To leverage the vast amount of real-world patient data with the tools developed to work with ontologies, an accurate mapping between the different systems used for annotating these datasets is necessary. In this paper we have analyzed how well aligned the International Classification of Diseases (ICD) is with the Human Phenotype Ontology (HPO). The rationale for moving from ICD codes to HPO terms is that, for example, genomics-based diagnostic pipelines rely on standardized phenotype and disease ontology terms as the input, such as terms from HPO. Only with a mapping from ICD-coded patient data to HPO can researchers accurately integrate and cross-analyze data originating from EHRs together with public research data beyond their study. For example, tools like PheRS, which measures the similarity between an individual's diagnosis codes and phenotypic features of known genetic disorders, also require mappings that link ICD codes to HPO terms, which most EHRs do not contain [21]. In such scenarios, unless the researcher is an expert in ontology, they will most likely turn to resources like the UMLS tables for these mappings, hence determining the coverage of ICD10-CM to HPO mappings with the UMLS table is imperative.

We set out to analyze the extent to which ICD codes—either specified in ICD10-CM (i.e., our "dictionary level" analysis) or used in our EHR data—were mapped to the widely used phenotype ontology HPO. In our analysis of the mapping coverage, we found that the proportion of terms mapped between ICD and HPO are low both at the dictionary level and in real-world EHR data. That said, the proportion of mapped terms in our EHR data is higher than that found at the dictionary level. However, to enable the integration and cross-analysis of clinical and experimental data at scale, we argue there is an urgent need to improve mappability across these ontologies, and to make such mappings available to the research community using such vehicles as the UMLS or other appropriate mechanisms. We also found that certain terms have much higher leverage in improving patient coverage and should be looked into first. For example, such ICD codes as those related to COVID-19 infection or other viral diseases, nicotine dependence, and anxiety, are each used to annotate over 10,000 patients in our dataset. It should also be cautioned that the proportion of ICD codes mapped to HPO is especially low for rarer conditions (e.g. ICD codes from the neoplasm category are considerably less mapped than codes from the nervous system category) and could implicate biases if the data is used as is after conversion.

Our study indicates that the coverage of ICD to HPO mappings has not increased over the years, in fact the opposite—coverage decreased. This likely occurred because ICD and HPO grew over time (i.e. ontology terms were added) while the mapping set between the two artifacts has seemingly not kept up with their evolution. In the last analysis of mapping coverage reported in the literature in 2014, 15% of HPO codes were mapped to ICD10-CM. Whereas currently, our analysis indicates that the percentage of HPO terms mapped to ICD10-CM decreased to only 11% in the 2022 version of the UMLS Metathesaurus. The UMLS website provides a similar statistic that roughly confirms our empirical finding, albeit UMLS's statistics are likely more up-to-date—according to the UMLS website, in November 2023 a total of 10.2% of HPO terms are mapped to ICD10-CM [22].

It should be noted that there is a mismatch in granularity between the two artifacts under analysis, with ICD10-CM being a more coarse-grained representation than HPO. So understandably the proportion of exact mappings between terms is rather low. In certain circumstances one could consider non-exact mappings between terms in these ontologies— for example, broad and narrow mappings between terms that take into account the hierarchical structures of ICD and HPO—which can still be useful for large data analytics or other data integration purposes. However, to our knowledge there are no off-the-shelf tools that can identify such broad or narrow mappings. Furthermore, currently there are no benchmarks or means by which we could ascertain the validity of such mappings, given the lack of publicly available broad/narrow mapping sets between ontologies that researchers could use. In future work we intend to broaden our analysis to consider other mapping resources (e.g., broad/narrow mappings, concept sets) and to determine the extent to which differences in granularity between ICD and HPO can be bridged using such resources.

The scientific value of having a go-to resource for expert-verified mappings (such as UMLS) must not be underestimated. Such a resource allows for turnkey conversion between metadata that describe scientific data using different ontologies or coding mechanisms. It is essential to have appropriate tooling to allow researchers to easily leverage these mappings for their purposes, without having to go through literature, broken links, tables with various formats, and so on. UMLS is a good starting source for mappings, but clearly there is a need for more comprehensive mapping between ontologies that can complement the current ones. When ontology mappings for entities of interest are not available in a trusted source like UMLS, researchers tend to rely on NLP tools—some of which are based on modern, neural network-based language models—to generate mappings between their codes or ontology terms of interest. Language models are statistical methods that learn the probability of word occurrences based on examples of text in order to make predictions involving those words, and they are used to power NLP applications. For example, the popular language model Bidirectional Encoder Representations from Transformers (BERT) generates numeric (vector) representations of strings, which can then be compared in vector space using measures such as cosine distance. A comparison between two such entities will yield a distance score that researchers can use to determine how similar the two entities are in their meaning. In a similar vein, Large Language Models (LLMs) can be helpful in generating tentative mappings between ontologies. While the effectiveness of such approaches have yet to be evaluated, their outputs will certainly require review by qualified human curators and should not be used out-of-the-box. The reproducibility of mappings produced using generative AI models also needs to be considered. While we observed in our survey that such automated approaches are often used in practice, the mappings that researchers generate using these tools and use in their studies are rarely made publicly available—and so other scientists cannot reuse them, or even begin to try to replicate the original experiments that were done using those mappings. Even when researchers share their mappings, they are rarely (if ever) contributed to some centralized repository or registry of mappings that the wider community can leverage, such as UMLS.

## Conclusion

The low number of direct mappings between ICD and HPO imply these are unlikely to be sufficient for fine-grained phenotyping using EHR data. Our results also show that ICD terms with lower usage are less well mapped, hence the UMLS mappings are less likely to cover any existing ICD codes that might be a component of a rare disease. At best, the current state of mapping resources allows researchers to analyze relatively small cohorts whose diagnoses fall under the popular ICD categories that have been mapped.

## Ethical Considerations

IRB approval was obtained at the BIDMC (#2020P000565). Participant informed consent was waived by each IRB because the study involved only retrospective data and no individually identifiable data were shared outside of each site's local study team.

## Data Availability

To replicate this analysis it is necessary to have (1) the UMLS Metathesaurus (version 2022AA) and (2) the Beth Israel Deaconess Medical Center (BIDMC) patient dataset:

(1) UMLS is freely available upon registering for a UMLS Terminology Services (UTS) account (https://uts.nlm.nih.gov/uts/signup-login). The UMLS Metathesaurus version 2022AA that we used in our study can be obtained from: https://tinyurl.com/m7976xcz.

(2) The BIDMC dataset contains confidential patient electronic health data and cannot be shared outside of the institution. Interested parties may contact authors for potential collaboration or contact BIDMC's IRB to request access to the data through: https://www.bidmc.org/research/research-and-academic-affairs/clinical-research-at-bidmc/committee-on-clinical-investigation-irb.

## Code Availability

The code to replicate the analysis (tables, figures, etc.) is available on GitHub at: https://github.com/hms-dbmi/ICDtoHPOmappings.

## Authors' Contributions

**Design and Conceptualization of study:** Amelia Tan, Rafael Gonçalves, Isaac Kohane
**Data Collection:** William Yuan, Gabriel Brat
**Data Analysis or Interpretation:** Amelia Tan, Rafael Gonçalves, Isaac Kohane
**Drafting and Revision of Manuscript:** Amelia Tan, Rafael Gonçalves, William Yuan, Gabriel Brat, Robert Gentleman, Isaac Kohane

## Declaration of interests and source of funding statements

Robert Gentleman owns shares or stocks or consults for: 23andMe, Scorpion Tx, BioMap, Myia Labs, Pheno.AI, intrECate.

The authors have no sources of funding to declare.

## Figure Legends

Figure 1. The proportion of ICD/diagnosis codes that are matched to a HPO term, unmatched to any HPO terms or others (do not have a corresponding ICD code in the UMLS dictionary). The proportions were calculated for common codes that are used in >1% of the cohort, infrequently used codes that are attributed to 0.1-1% of the cohort and rare codes that are assigned to <0.1% of the patient cohort. These were calculated separately for the (A) admitted patients, (B) ICU patients, and (C) outpatients.

Figure 2. Patient coverage of mapped and unmapped terms across different ICD categories. The left column specifies codes that are used in less than 100 patients, while the right column specifies codes that are used in more than 100 patients in our dataset. The figure depicts In yellow the frequency of usage of ICD codes that have an HPO mapping, in green the usage of ICD codes that do not have an HPO mapping, and in purple the codes in the "others" category which are in the ICD code list of our EHR dataset, but which did not match existing ICD10-CM codes. The numbers on the right of each column specify the number of times that codes from that category are used.